# Lightweight Cryptography for IoT: A State-of-the-Art

**VISHAL A. THAKOR[1], M.A.RAZZAQUE[2]\*, and MUHAMMAD R. A. KHANDAKER[3],**

[1]School of Computing, Engineering Digital Technologies, Teesside University, UK (e-mail: v.thakor@tees.ac.uk)
[2]School of Computing, Engineering Digital Technologies, Teesside University, UK (e-mail: m.razzaque@tees.ac.uk) \*Corresponding author
[3]School of Engineering and Physical Sciences, Heriot-Watt University, UK (email: m.khandaker@hw.ac.uk)

**ABSTRACT:** With the emergence of 5G, Internet of Things (IoT) has become a center of attraction for almost all industries due to its wide range of applications from various domains. The explosive growth of industrial control processes and the industrial IoT, imposes unprecedented vulnerability to cyber threats in critical infrastructure through the interconnected systems. This new security threats could be minimized by lightweight cryptography, a sub-branch of cryptography, especially derived for resource-constrained devices such as RFID tags, smart cards, wireless sensors, etc. More than four dozens of lightweight cryptography algorithms have been proposed, designed for specific application(s). These algorithms exhibit diverse hardware and software performances in different circumstances. This paper presents the performance comparison along with their reported cryptanalysis, mainly for lightweight block ciphers, and further shows new research directions to develop novel algorithms with right balance of cost, performance and security characteristics.

**INDEX TERMS:** Internet of Things (IoT), Lightweight, Cryptography, Cryptanalysis

## I. INTRODUCTION

### A. IOT OVERVIEW

Internet of Things (IoT) has already become a dominant research era because of it's applications in various domains such as smart transportation & logistics, smart healthcare, smart environment, smart infrastructure (smart cities, smart homes, smart offices, smart malls, industry 4.0), smart agriculture and many more (Figure 1). Many researchers and industry experts have given various definitions of IoT de- pending on their applications and implementation area, but in simple words, IoT is a network of connected objects, each with a unique identification, able to collect and exchange data over the Internet with or without human interaction [1]–[5]. Such connected objects (industrial controllers, RFID tags, sensor nodes, smart cards, home appliances etc.) are becoming more common and will flood the market with the emerge of IoT [6], leading an enormous data exchange rate amongst [7].

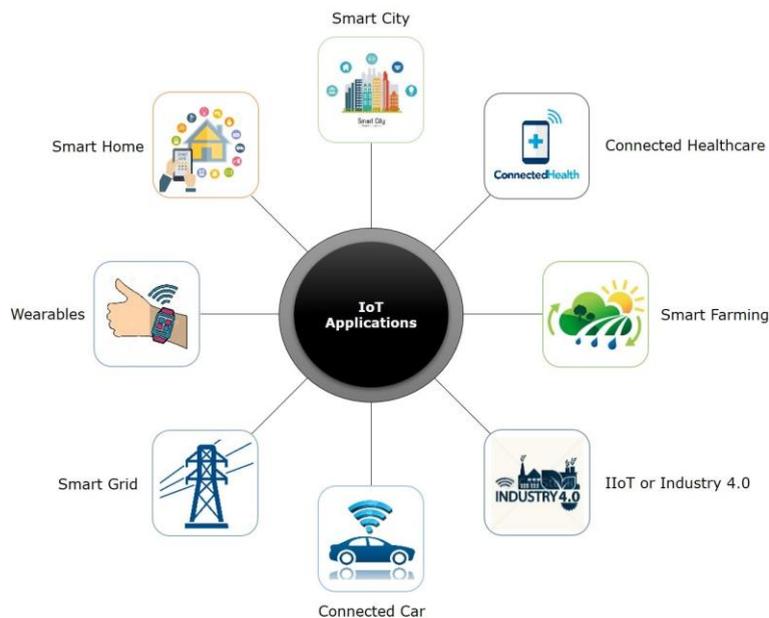

**Figure 1.** IoT Applications

## B. IOT SECURITY CONCERNS

When billions of smart devices working in a diverse set of platforms are connected to the Internet, especially when shifting from desktop computers to small devices, they bring a wide range of new and unprecedented challenges to their owners or users [6] such as security & privacy, interoperability, longevity & support, technologies and many more [8]. Also, IoT devices are easily accessible and ex-posed to many security attacks [9] as they interact directly with the physical world to collect confidential data or to control physical environment variables, which makes them an attractive target for attackers [10]. All these circumstances make cybersecurity as a major challenge in IoT devices with demands of confidentiality, data integrity, authentication & authorization, availability, privacy & regulation standards and regular system updates [8]. In this scenario, cryptography could be one of the effective measures to guarantee confidentiality, integrity and authentication & authorization of the traversing data through IoT devices [7]. Cryptography could also be a solution to secure the stored or traversing data over the network. However, traditional PC based cryptographic solutions are not suitable for most IoT devices, including sensor and RFID tags, due to their resource limitations. A lighter version of these solutions can address this challenge. Generally, the lighter versions of computation-cryptography are known as lightweight cryptography (LWC).

Recently, a number of algorithms have been proposed for LWC. In addition, many works have revealed the security attacks on particular LWC algorithm(s) [11]–[27]. A number of published papers have done a fair comparison of hard- ware and/or software implementations of these algorithms on different platforms as well as in different circumstances [9], [28]–[35]. Most of these works have considered the algorithms which are applicable in certain domains or suit- able for certain applications. However, a holistic view of the proposed LWC algorithms in terms of their hardware- software performances along with cryptanalysis is missing in these works. Authors in [36] have reviewed a list of different LWC algorithms (still missing some key algorithms, e.g., Midori [37]) with their performances on different platforms but missing an inclusive view on their applications and key demands of lightweightness. With a unique aspect in this paper, we focus on key characteristics of LWC algorithms (Table 1) defined by leading research groups in the fields of cryptography and have evaluated them in terms of hardware and software metrics. In addition, demonstrating various IoT applications in real world along with their lightweight key requirements and their best suite LWC options is another motivation behind the work.

Considering the importance of IoT security, this article takes a holistic view of symmetric key lightweight cryptography algorithms and i) identifies the key characteristics of LWC, and the requirements of LWC (Section II), ii) based on the identified requirements, presents a comprehensive review of the existing LWC algorithms focusing on state- of-the-art research (Section III), iii) hardware and software performances analysis (Section IV) and cryptanalysis (Section V) of the algorithms, and finally iv) outlines open re- search challenges, recommending future research directions (Section VI).

## II. IOT AND LIGHTWEIGHT CRYPTOGRAPHY

The key challenges of implementing conventional cryptography in IoT devices are as follows [38]:
- Limited memory (registers, RAM, ROM)
- Reduced computing power
- Small physical area to implement the assembly
- Low battery power (or no battery)
- Real-time response

Most of the IoT devices (such as RFIDs and sensors) are small in size and are equipped with limited resources such as small memory (RAM, ROM) to store and to run the application, low computing power to process the data, limited battery power (or no battery in case of passive RFID tags) [6], small physical area to fit-in the assembly [6] [38]. Moreover, most of the IoT devices deal with the real time application where quick and accurate response with essential security using available resources is a challenging task [39] [40]. IoT device designers face several risks and challenges, including energy capacity [41], and data security [9].

In these circumstances, if conventional cryptography standards are applied to IoT devices (mainly RFIDs and sensors), their performance may not be acceptable [6]. The above issues with conventional cryptography are very well addressed by its sub-discipline, lightweight cryptography, by introducing lightweight features such as small memory, small processing power, low power consumption, real time response even with resource constrained devices [6].

Another important aspect of lightweight cryptography is that it is not just limited to a particular device that drives the need for lightweight cryptography (RFID tags, sensors, etc.), but readily applicable to other devices rich in resources that it directly or indirectly interacts with (such as cellphones, tablets, PCs, servers, etc.) [6].

The three main characteristics of Lightweight cryptography algorithms and their offerings are listed in Table 1 [9] [38]:

| Characteristics | | What LWC can offer? |
|---|---|---|
| Physical (Cost) | Physical Area(GEs, logic blocks) | Smaller block sizes (64-bit or less) |
| | Memory (registers, RAM, ROM) | Smaller key size (80-bit or less) |
| | Battery power (energy consumption) | Simple round logic based on simple computations |
| Performance | Computing Power (latency, throughput) | Simple key scheduling |
| Security | Minimum security strength (bits) | Strong Structure (like SPN or FNS) |
| | Attack models (related key, multi- keys) | |
| | Side channel attack | |

**Table 1.** LWC Characteristics

### A. HARDWARE AND SOFTWARE PERFORMANCE METRICS

The hardware specific requirements of any digital device are typically measured in terms of memory requirements, gate area, latency, throughput, and power and energy consumption as follows:

**Memory requirements:** Generally, measured in KB [36]. RAM (Random Access Memory) required to store intermediate values that can be used in computations and ROM (Read Only Memory) required to store the program code (algorithm), and fixed data, such as S-boxes or hardcoded round key [6].

**Gate Area:** The size of the physical area required to implement/run the algorithm on a circuit can be defined in terms of logic blocks for Field-Programmable Gate Arrays (FPGAs), or by Gate Equivalents (GEs) for ASIC implementations (one GE equivalent to two-input NAND gate) [6]. It is measured in $\mu m2$. Normally, a low-cost RFID tag are capable of accommodating around 1000 to 10,000 gates, out of which only 200 to 2000 may be available for security reasons [42].

**Latency:** It is measure of time between the initial request of an operation and producing the output (time between production of ciphertext from plaintext) [6].

**Throughput:** Throughput in hardware can be measured in terms of plain text processed per time unit (bits per second) at 100 *KHz* frequency, whereas in software, it is average amount of plaintext processed per CPU clock cycle at 4 *MHz* frequency [43].

**Power and energy consumption:** The amount of power required by the circuit (consumed by the hardware or soft- ware implementation) is measured in $\mu W$. Energy consumption per bit is calculated by the formula [36]:

$$Energy\ [\mu J] = (Latency\ [cycles/block] * Power\ [\mu W])/blocksize\ [bits]$$

Here, latency is the number of clock cycles that are required to encrypt a block

**Efficiency:** Gives the trade-off between performance and implementation size. For hardware, it can be calculated as follows [36]:

$$Hardware\ Efficiency = Throughput\ [Kbps]/Complexity\ [KGE]$$

Here, complexity is the value of the physical area in *KGE*.

Similarly, **software efficiency** can be calculated as follows [36]:

$$Software\ Efficiency = Throughput\ [Kbps]/CodeSize\ [KB]$$

Here, code size is the size of the executable code in *KB*.

## B. WHY SYMMETRIC BLOCK CIPHER?

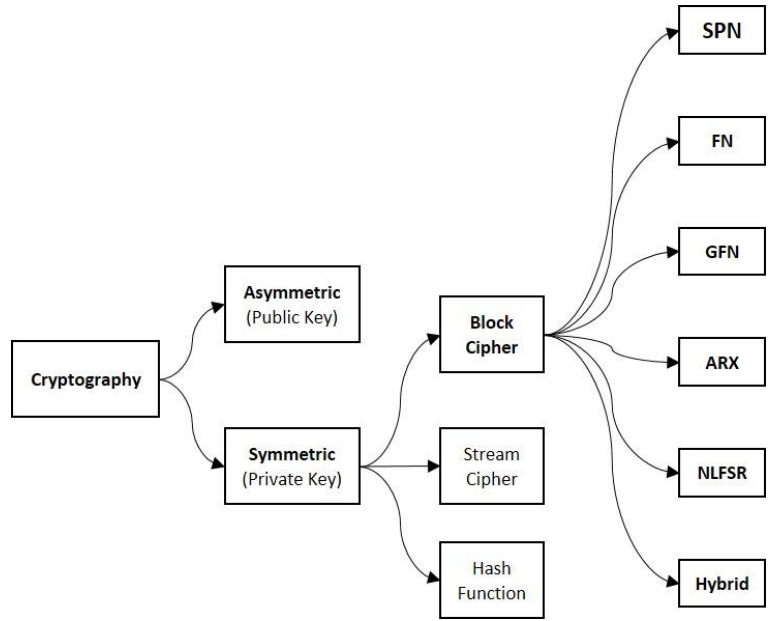

**Figure 2.** Classification of Cryptography

Cryptography can be classified into two main branches, symmetric cipher and asymmetric cipher (Figure 2). Symmetric cipher uses single key for both encryption and decryption of the data, whereas asymmetric cipher uses two keys, public key for encryption and private key for decryption [44]. Symmetric key encryption is extremely secure and relatively fast, only disadvantage is sharing of the keys between the communicating parties [28]. In addition, it assures confidentiality and integrity of data but do not guarantee authentication. Asymmetric encryption ensures confidentiality and integrity by allowing sender to use public key of the receiver where receiver use his private key to decrypt the cipher. At the same time, it guarantees authentication by digital signature, which allows sender to use his/her private key too to encrypt the data further and receiver decrypt it by using sender's public key [44]. The only disadvantage of asymmetric encryption is its large key which increases the complexity and slowdown the process [28].

In block cipher, both encryption and decryption happen on a fixed size block (64 bits or more) at a time whereas stream cipher processes whole message byte by byte (8 bits at a time). In addition, stream cipher uses only confusion property (to make relationship as complex as possible using substitution between the cipher text and the key [31] [44]) whereas block cipher uses both confusion and diffusion (to dissipate statistical structure of plaintext over bulk of cipher text using permutation [31] [44]), two fundamental properties of cryptography introduced by Claude Shannon [31] [36], to strengthen the cipher with simple designing compare to the stream one. Reversing encrypted text is hard in block cipher whereas stream cipher uses XOR for the encryption which can be easily reversed to the plain text. In contrary, Hash function is a mathematical algorithm that maps data of arbitrary size to a bit string of a fixed length hash (short). It is infeasible to invert, one-way function.

For the above reasons, block cipher is preferred in resource-constrained IoT devices over stream cipher, and this paper is concentrating on block cipher. A symmetric lightweight block cipher uses one of the following structure:

- Substitution-Permutation Network (SPN)
- Feistel Network (FN)
- General Feistel Network (GFN)
- Add-Rotate-XOR (ARX)
- NonLinear-Feedback Shift Register (NLFSR)
- Hybrid

**Substitution-Permutation network (SPN)** processes data through a series of substitution (S-box) and permutation (table) by altering the data and finally formulating them for the next round. A **Feistel network (FN)** is a multi-round cipher that divides the input block into two parts and operates only on a half (diffusion) in each round of encryption or decryption. Between rounds, the left and right halves of the block are swapped. The **generalized Feistel network (GFN)** is a generalized form of the classical Feistel cipher. In GFN, input block is split into two or more sub-blocks and applies a (classical) Feistel transformation for every two sub blocks, and then performs a cyclic shift relevant to number of sub blocks [45]. **ARX** performs encryption-decryption using addition, rotation and XOR functions without making any use of S-box. Implementation of ARX is fast and compact but limits in security properties compared to SPN and Feistel ciphers. **Nonlinear feedback shift register (NLFSR)** can be applied in both stream cipher and block cipher designs. It utilizes the building blocks of stream ciphers whose current state is a nonlinear feedback function of its previous state [16]. **Hybrid** cipher combine the any three types (SPN, FN, GFN, ARX, NLFSR) to improve specific characteristic (for example, throughput, energy, GE, etc.) based on its application requirements or even could mix of block and stream cipher.

Out of these structures, SPN and FN are the most popular choice due to their flexibility to implement the structure based on application requirements [36]. Feistel structures can be implemented in low average power hardware, as a round function is applied to only one half of the state [37]. On contrary, Feistel structures apply non-linearity in just one half of the state in each round, maintaining safety margins usually requires more round function compared to SPN structures. When there is choice between fewer SPN function rounds and higher Feistel function rounds with the same level of security and similar energy costs, SPN function could be a smarter choice [37].

## III. EXISTING WORKS

More than fifty symmetric LWC algorithms are proposed by various academia, proprietaries and government bodies with focus on reducing cost (memory, processing power, physical area (GE), energy consumption) and enhanced hardware and software performance (latency, throughput). However, many of them do not concentrate on security attacks explicitly [40]. The structure wise categorization of these algorithms are summarized in Table 2. In the following subsections, we present an overview of these algorithms within each structural category of LWC.

| Structure Type | Algorithms |
|---|---|
| SPN | AES, PRESENT, RECTANGLE, MIDORI, mCrypton, NOEKEON, ICEBERG, PUFFIN-2, PRINCE, PRIDE, PRINT, Klein, LED, PICARO, ZORRO, I-PRESENT, EPCBC |
| FN | DESL/DESXL, TEA/XTEA/XXTEA, Camellia, SIMON, SEA, KASUMI, MIBS, LBlock, ITUbee, FeW, GOST, Robin, Fantomas |
| GFN | CLEFIA, PICCOLO, TWIS, TWINE, HISEC |
| ARX | SPECK, IDEA, HIGHT, BEST-1, LEA |
| NLFSR | KeeLoq, KATAN/KTANTAN, Halka |
| Hybrid | Hummingbird, Hummingbird-2, PRESENT-GRP (SPN+GRP (Group Permutation)) |

Table 2. Structure wise LWC algorithms

### A. SUBSTITUTION PERMUTATION NETWORK (SPN)

**AES** [46] is a classic example of SPN based algorithm, standardized by NIST (National Institute of Standards and Technology) and takes on 128-bits block with variable key sizes of 128, 192, 256 bits [47]. The minimum GE requirement recorded for AES is around 2400 GEs (23% smaller than the usual one) [47], which is still heavy for some small scale real-time applications [31]. It shows comparatively efficient performance when supplied with additional resources [34].

Another, most hardware and software efficient and ISO/IEC(29192-2P:2012) approved algorithm is **PRESENT** which also uses Substitution-Permutation network based on 64-bit block size with two key variants; 80-bit and 128- bit key [48]. It's ultra-lightweight version requires approx. 1000 GEs (encryption only), where it takes 2520-3010 GEs to provide adequate level of security [31]. It is a hardware efficient algorithm and uses 4-bit S-boxes (substitution layer replaces eight S-boxes with single S-box) whereas it takes large cycles in software (permutation layer) which demands an improved version of this [28] [31] [36] [48] [49].

**RECTANGLE** is an ultra-lightweight block cipher that can be used with various application. With little changes in SPN structure, the number of round are reduced to 25 (compared to 31 rounds in PRESENT) to meet with the competitive environment [49].

**TWINE** achieves good overall status as PRESENT as well as overcomes many of its implementation issues. It operates on 64-bits block using 80-bits and 128-bits key sizes. It requires around 2000 GE and circuit size per throughput is more than twice compare to AES. In speed comparison, when 1KB or more ROM is available, AES is faster than TWINE but when only 512bytes of ROM is available, AES can't be implemented and works 250% faster than PRESENT.

**Midori** was designed with focus on low/tight energy budget, for instance, medical implants. It comes with two different versions, Midori64 and Midori128. Both of these use the same key size 128-bit on different block size 64-bit and 128-bit with 16 and 20 rounds, respectively [37] [50].

**mCrypton** (miniature of Crypton) [51] is a compact edition of Crypton [52], designed for low-cost RFID tags and sensors and exhibits low-power and compact implementations in both hardware and software. It uses substitution- permutation network (SPN) on 64-bit block through 13 rounds and variable key sizes 64-bit, 96-bit and 128-bit. A related key rectangle attack was reported on 8th round of mCrypton [53].

**NOEKEON** [54] is a block cipher uses substitution- permutation network (SPN) with a block and key length of 128 bits through 16 identical rounds. A related-key crypt- analysis was presented in [55] and as a result the cipher was rejected by the NESSIE project.

**ICEBERG [56]** is designed for re-configurable hardware implementations as it allows changing the key at every clock cycle without compromising performance by deriving round keys on-the-fly. It uses 128-bit keys with 64-bit blocks through 16 rounds requires 5800 gates for 400 Kbps of throughput [57]. ICEBERG suffers from differential key attack on 8th round [58].

**PUFFIN-2** [59] is based on PUFFIN (2303GE) [60], a serialized SPN architecture and supports the 80-bit key and 64-bit block through 34 rounds. It requires only 1083 GEs for both encryption and decryption. Differential cryptanalysis is the weakness of PUFFIN-2 [61].

**PRINCE** [62] is both hardware and software [63] efficient lightweight algorithm which uses 128-bit keys with 64-bit blocks through 12 rounds. The most compact hardware implementation requires 2953GE for 533.3 Kbps of throughput and has low energy consumption (5.53 J/ bit) [64]. Attacks on the reduced 6-round and the full 12-round [65] [66], truncated differential cryptanalysis on 7-round [67] were reported on PRINCE.

**PRIDE** [63] uses 128-bit keys with 64-bit blocks through 20 rounds and exhibits low latency and energy consumption. PRIDE does not claim any resistance against related- key attacks and generic time-memory trade-offs are possible against PRIDE.

**PRINT** [68] is a domain specific cipher designed for two applications: PRINT-48 for IC-printing applications which make use of 80-bit key and 48-bit block through 48 rounds (402GE) and PRINT-96 for EPC encryption which uses 160- bit key and 96-bit block through 96 rounds (726GE). It uses 3-bit operations where odd number of bit operation is not feasible, actual deployment of the algorithm is not ready yet. Related-key cryptanalysis on the full round cipher were presented in [69].

**Klein** [70] uses 64-bit blocks with 64-bits, 80-bits and 96-bit keys through 12 (1220 GEs), 16 (1478 GEs), and 20 (1528 GEs) rounds respectively. It targets legacy sensor platforms. It was designed with focus on software implementation, mainly for sensors. The chosen-plaintext key-recovery attacks [71] and an asymmetric biclique attack [72] are the weaknesses of Klein algorithm.

**LED** (Lightweight Encryption Device) [73] is designed with a novel concept of no key scheduling process, the PRESENT S-box, the row-wise processing as described in lightweight AES [47] and the mix column approach of the hash function PHOTON [74] to achieve small hardware footprint and reasonable software performance. It uses 64- bit (966 GE), 80-bit (1040 GE), 96-bit (1116 GE) and 128-bit (1265 GE) keys with 64-bit blocks through 32 and 48 rounds. This approach reduces the chip area but increase the security risk at the same time, like related key attacks [75]. Biclique cryptanalysis on reduced round cipher [24] and differential fault analysis based on Super-Sbox techniques [76] was obtained on LED.

**PICARO** [77] is a novel cipher having a good trade-off between efficiency, conventional security and masking efficiency (by an adequate choice of S-box). It has 4 different masking levels compare to existing ciphers, like AES and the hardware implementation is faster than the corresponding AES too. It uses 128-bit key through 12 rounds and shows high resistance to side channel attacks.

**Zorro** [78], based on AES, is suitable for embedded systems and more efficient than PICARO. It uses same block and key size (128-bit) through 24 rounds. However, practical invariant subspace attacks are presented in [79].

**EPCBC** (Electronic Product Code Block Cipher) [80] is a lightweight cipher, inspired by PRESENT, supports 96- bit key with 48-bit and 96-bit block through 32 rounds. The most compact version needs 1008GE. The security level is improved by optimizing key scheduling procedure that immunes it against related key differential attacks. Algebraic attack up to 5 rounds of EPCBC-96 and up to 8 rounds of EPCBC-48 using known plaintexts and weak keys for EPCBC-96 with up to 7 rounds were reported in [19].

**I-PRESENT** [81] is an involutivity version of PRESENT inspired from PRINCE and NOEKEON. It uses the same key and block sizes through 30 rounds with two additional 4x4 S- boxes 16 times. The most compact hardware implementation requires about 2769 GE (encryption and decryption).

### B. FEISTEL NETWORK (FN)

The lightweight DES (Data Encryption Standard) is known as **DESL**. It uses 64-bit block with 56-bit key by performing 16 rounds (same as DES) by replacing eight original S-boxes with a single one [82] as well as the multiplexer [83], requiring 1850 GEs (20% compact chip design compare to DES (2310 GEs)) [83]. DESL also discards the initial and final permutation of DES to make it more lighter [84]. **DESXL** is lightweight version of DES with key whitening feature to strengthen the cipher. It performs the same number of cycles and uses same block size as DESL but larger key, 184-bit (k=56, k1=64, k2=64) [84] with 2170 GEs requirements [83]. DESL resists common attacks such as linear and differential cryptanalysis and the Davies-Murphy attack but demands more resources compare to other lightweight competitors [83] [84].

**Tiny Encryption Algorithm (TEA)** is suitable for very small, computationally weak and low cost hardware [85]. It uses Feistel Structure (FN) and performs 32 rounds on a 64-bit block using 128-bit key [86] with equivalent gate requirements of 3872 [87]. It's simple key scheduling is vulnerable to brute force attack [88] [89]. Another limitation of TEA structure is it's three equivalent keys for decryption which makes it vulnerable to the attackers [88]. The improved version of TEA is **eXtended TEA (XTEA)** which uses the same combination of block and key size but with increased number of rounds (64 rounds) with 3490 GE [90] requirements. It offers more complex key scheduling with little change in Shift, XOR and addition operations [91]. XTEA was victim of related-key rectangle attack (on 36 rounds) [91] and further modified with **XXTEA** [92]. A chosen-plaintext attack based on differential analysis against the full-round cipher was reported in [93].

**Camellia** [94] was designed by Nippon Telegraph and Telephone Corporation and Mitsubishi Electric Corporation and is recognized by ISO/IEC, IETF, projects NESSIE, CRYPTREC and Japan's new e-Government Recommended Ciphers List. It uses the same block and key sizes as AES through 18 and 24 rounds and offers similar level of security as AES. It is known for its fast software implementations as the hardware implementation [95] requires 6511 GE. Cache timing attacks in software implementations were presented in [20].

**SIMON** [96], designed by National Security Agency (NSA), is known for its optimal performance in hardware. It uses various key size 64-bit, 72-bit, 96-bit, 128-bit, 144-bit, 192-bit, 256-bit and block size 32-bit, 48-bit, 64-bit, 96-bit, 128-bit through 32, 36, 42, 44, 52, 54, 68, 69, 72 rounds [96]. The most compact version requires 763GE for execution [96]. Various attacks such as differential and impossible differential attacks [25], differential fault attack [97], Cube and dynamic cube attacks [26], recover the full key in a practical time complexity [26] were reported on reduced- round versions of SIMON [98].

**SEA** (Scalable Encryption Algorithm) [99] is a parametric block cipher designed for resource constrained such as sensor networks, RFIDs, etc., [100] with a goal to meet low memory requirements, small code size and a limited instruction set with the ability to derive keys on-the-fly [99]. It uses 96- bit key on two recommended block size 96-bit and 8-bit with requirement of 3758GE [100] for the most lightweight hardware version. The optimized software implementation requires 426 bytes of code and 41604 cycles for encryption on 8-bit micro-controllers [101].

**KASUMI** [102] uses 128-bit keys with 64-bit blocks through 8 rounds with requirements of 3437GE [103], mainly designed for GSM, UMTS and GPRS systems. KASUMI is victim of differential-based related-key attack [13] and a single-key attack on 6th round [14].

**MIBS** [104] uses 64-bit blocks with 64-bit (1396 GE) and 80-bit (1530GE) keys through 32 rounds. It uses Feistel structure with an SPN round function (S-box of mCrypton [51]) and PRESENT based key scheduling.

Various attacks, such as linear attacks (up to 18 rounds), first ciphertext-only attacks (on 13 rounds), differential analysis (up to 14 rounds) and impossible-differential attacks (up to 12 rounds) were reported in [105].

**LBlock** [106] is an ultra-lightweight cipher, uses 80-bit keys and 64-bit blocks through 32 rounds. The most compact hardware implementation requires 1320GE for 200 Kbps throughput whereas 3955 clock cycles are taken by most efficient software implementation to encrypt a single block (on 8-bit microcontroller). Many security attacks reported by researchers against LBlock are saturation attack on 22-round [107], biclique cryptanalysis against the full round [108], improved differential fault attacks [21] and round addition differential fault analysis [22].

The originally developed by the government of the Soviet Union (1989), the lightweight version of **GOST** uses 256-bit key on 64-bit block size with 32 rounds requiring 651GE by adapting S-Box from PRESENT [109]. A theoretical meet-in-the-middle attack [110] and a single-key attack [111] were analyzed on GOST.

**ITUbee** [112] is a software efficient cipher that requires 586 bytes of code and 2937 cycles for encryption (most com- pact version). It uses 80-bit keys with 80-bit blocks. Here, key scheduling is replaced by round-dependent constants to reduce software overload. Self-similarity cryptanalysis on 8th round is performed in [113].

**FeW** [114] supports 64-bit block with two key variants, 80-bit and 128-bit keys over 32 rounds. It makes use of S- box of Humminbird-2 and follows the key expansion process from PRESENT. There no cryptanalytic attack found on FeW [114].

### C. GENERALISED FEISTEL NETWORK (GFN)

SONY corporation introduced **CLEFIA**, standardized by NIST in 2007, which is based on Feistel structure, uses 128-bit block with choice of 128, 192, 256 bit key through 18, 22, 26 round, respectively [115] [116]. It shows high performance and strong immunity against various attacks [36] [115] [117] [118] with comparative high cost as the most compact version requires 2488 GE (encryption only) for 128- bit key [116]. CLEFIA has two confusion and two diffusion properties which makes it stronger against different attacks but at the same time these demands higher memory and limits its use in ultra-small applications [31].

**Piccolo** [119] is another ultra-lightweight cryptography algorithm suitable for extremely constrained environments such as RFID tags and sensor nodes. It processes 64-bit block cipher with two key variants, 80-bit and 128-bit keys through 25 and 31 rounds. The most compact hardware implementation (80-bit key) requires 432 GE with only 60 additional GE to support the decryption. Biclique attacks on full round [24] [120], impossible differential cryptanalysis on reduced round versions [121] were reported on Piccolo.

**TWIS** [122] derived from CLEFIA, uses same key and block size (128-bit) through 10 rounds. It is victim of differential distinguisher with probability 1 [123].

**TWINE** [107], LBlock based, is a 36 round block cipher with a 64-bit state and two key sizes, 80-bit and 128-bit key. The most compact hardware implementation requires 1866GE. TWINE uses nibble permutation and a single S-box while LBlock uses bit permutation (for key scheduling) and ten S-Boxes. The meet-in-the-middle attacks was reported on simplified key-scheduling in [124].

**HISEC** [125] uses 80-bit keys with 64-bit blocks through 15 rounds (1695GE). It shows good resistance against different attacks and the characteristics are more like to PRESENT except bit-permutation.

### D. ADD-ROTATE-XOR (ARX)

**SPECK** [96] is designed by National Security Agency (NSA) along with SIMON. It is software-oriented cipher supports the same key and block sizes as SIMON through 22, 23, 26, 27, 28, 29, 32, 33 and 34 rounds. The most compact hardware implementation recorded uses 48-bit block with 96-bit key with requirements of 884 GE whereas the most efficient software implementation requires 599 cycles with 186-byte of ROM for 64-bit block with 128-bit key [96]. An attack on reduced versions [126], differential fault attacks [97] are documented on SPECK.

**IDEA** (International Data Encryption Algorithm) [127], designed by Lai and Massey, uses 128-bit keys with 64- bit blocks through 8.5 rounds, mainly used for high-speed networks [128]. It uses 16-bit unsigned integer and perform data operations such as XOR, addition and modular multiplication without using S-box or P-box. IDEA performs well in embedded software (PGP v2.0). Its most efficient software implementation requires 596 bytes of code for 94.8 Kbps throughput [129]. It suffers from narrow-biclique attacks [12].

**HIGHT** (High security and lightweight) [130], an ultra- lightweight algorithm, processes 64-bit block with 128-bit key over 32 rounds using compact round function (no S- boxes) and simple computational operations. The most com- pact version acquires 2608 GE for 188 Kbps throughput [131]. Various attacks such as an impossible differential attack on 26-round [132], a related-key attack on full round [133], biclique cryptanalysis on full round version [120] and zero-correlation attacks on the 26- and 27-round cipher [11] were present on HIGHT.

**BEST-1** (Better Encryption Security Technique-1) [134], an ultra-lightweight cipher, targets wireless Sensor network and RFID tags. It operates on 64-bit block with 128-bit key through 12 rounds on 8-bit processors. The core functions of BEST-1 are mod 28 addition and subtraction, bitwise shift and XOR. It requires 2200 GE and achieves 265.7 Mbps at 80 MHz frequency.

**LEA** (Lightweight Block Encryption) [135], designed by the Electronics and Telecommunication Research Institute of Korea for 32-bit common processor. It uses 128-bit blocks with 128-bit, 192-bit and 256-bit keys through 24, 28, and 32 rounds respectively. LEA, software-oriented cipher, requires 590 bytes of ROM and 32 bytes of RAM for 326.94 cycles per byte speed on ARM platform. The most compact version requires 3826 GE for 76.19 Mbps throughput [136]. The 128- bit key is retrieved by attacking the first round using power analysis on LEA [23].

### E. NONLINEAR-FEEDBACK SHIFT REGISTER (NLFSR)

**KeeLoq** [18] uses 64-bit keys with 32-bit blocks through 528 rounds. It is widely used in various wireless authentication applications such as remote key less entry systems in cars [137]. It was created by Gideon Kuhn in the 80's but the first cryptanalysis of KeeLoq was published by Bogdanov [138] in February 2007. A slide attack and a novel meet-in-the-middle attack (direct algebraic attacks can break up to 128 rounds of KeeLoq) were reported in [137]–[139].

**KATAN/KTANTAN** [140], inspired from KeeLoq, cipher family applies 80-bit key on various block size (32-bit, 48-bit and 64-bit) through 254 rounds. They have small hardware footprint (KATAN 802GE and KTANTAN 462 GE), mainly designed for RFID tags and sensor networks, follow a linear structure (LFSR) instead of NLFSR of KeeLoq. KATAN has a very simple key scheduling compare to KeeLoq whereas KTANTAN exhibits no key generation operations (reduce GE requirement). KTANTAN is only used for devices where the key is initialized once and remain unchanged. KTANTAN- 48 (588 GE) is more appropriate for RFID tags. In soft- ware, both shows poor performance (low throughput and high energy consumption) due to overuse of bit manipula- tion [101]. Multidimensional meet-in-the-middle attacks on reduced round KATAN [141] and related-key attacks which recover the full 80-bit key of KTANTAN [15] were listed.

**Halka** [142] performs well on both hardware and software. It uses 80-bit keys with 64-bit blocks through 24 rounds. The multiplicative inverse based 8-bit S-boxes using LFSR makes it more secure than PRESENT which takes only 138GE (7% less GE than PRESENT) [142]. Also, the software performance is 3 times more efficient than PRESENT [142].

### F. HYBRID

**Hummingbird** [143] is an ultra-lightweight cipher which uses a hybrid structure (block and stream cipher), with 256- bit key on 16-bit block through 20 rounds. It was vulnerable to several attacks [144].

**Hummingbird-2** [145], designed for low end microcon- trollers, uses 128-bit keys and 64-bit Initialization Vectors which is efficient in both hardware and software. It also satisfies ISO 18000-6C protocol. It gives better performance compare to PRESENT (on 4-bit microcontrollers) but have few drawbacks: 1) Initialization is necessary before encryp- tion (or decryption) due to its stream property 2) Different encryption and decryption functions and due to that full version is 70% heavier than only encryption. Moreover, its performance degrades while processing small messages. Also, related key attack is present in [146].

**PRESENT-GRP** [31] processes 128-bit key with 64-bit blocks over 31 rounds. It makes use of SPN of PRESENT with a GRP (group operation) structure for higher confusion properties by replacing P-box of the original PRESENT. The hardware implementation of PRESENT (1884 GE) is slightly better than PRESENT-GRP (2125 GE). Similarly, PRESENT is more efficient than PRESENT-GRP in software implementation too.

## IV. HARDWARE AND SOFTWARE PERFORMANCE COMPARISON

Various experiments have been carried out by many researchers using different platforms such as NXP [31], AVR [147], ARM [31] micro-controllers to evaluate the performance of the popular lightweight cryptography algorithms [31] [34] [36] [40] [47] [83] [147] [148]. During these experiments, various characteristics such as area (GE), logic process (µm), power consumption (µW), through- put, RAM/ROM (bytes) requirements, latency (cycle/block), etc. have been compared for different lightweight cryptography algorithms in different circumstances (file types (C/C++, Java, Python), message size, etc.). Table 3 summarizes the hardware and software performance aforementioned lightweight cryptography algorithms. Figure 3 reveals the hardware and software efficiency of these LWC algorithms. According to the graph, software efficiency competition is won by SPECK, followed by SIMON and then PRIDE. Also, IDEA, ITUbee, LEA and AES show better software efficiency compare to the other LWC algorithms. In terms of hardware efficiency, Midori, Piccolo, GOST, PRINT, PRINCE, Rectangle, PRESENT, MIBS, LBlock, TWINE, and HIGHT are the principal competitors. In addition, Figure 4 and 5 respectively visualize the hardware and software performances (metrics wise) of these algorithms. Midori shows the lowest energy consumption (1.61µJ/bit), followed by Piccolo, PRINCE, TWINE and RECTANGLE. SPECK has the lowest latency rate (408 cycles/block) as well as smallest memory requirement (134 bytes), followed by the SIMON (594 cycles/block and 170 bytes). In general, there are two key observations from the study, (1) simple round leads to the hardware efficiency and (2) less number of rounds along with simplicity offer software efficiency.

| LWC Algorithm | Hardware Implementation | | | | | | | | Software Implementation | | | | | | | |
|---|---|---|---|---|---|---|---|---|---|---|---|---|---|---|---|---|
| | Key Size | Block Size | Tech (µm) | Area (GEs) | Power (µW) | Energy (µJ/bit) | Throughput @100KHz (Kbps) | Hardware Efficiency (Kbps/KGE) | Key Size | Block Size | ROM (byte) | RAM (byte) | Latency (Cycles/block) | Energy (µJ/bit) | Throughput @4MHz (Kbps) | Software Efficiency (Kbps/KB) |
| AES | 128 | 128 | 0.13 | 2400 | 2.4 | 42.38 | 56.64 | 23.6 | 128 | 128 | 918 | 0 | 4192 | 16.7 | 122 | 132.9 |
| PRESENT | 80 | 64 | 0.18 | 1570 | 2.35 | 11.77 | 200 | 127.38 | 128 | 64 | 660 | 0 | 10792 | 43.1 | 23.7 | 35.91 |
| RECTANGLE | 80 | 64 | 0.13 | 1467 | 1.46 | 5.96 | 246 | 167.68 | - | - | - | - | - | - | - | - |
| MIDORI | 128 | 64 | 0.09 | 1542 | 60.6 | 1.61 | 400 | 259.4 | - | - | - | - | - | - | - | - |
| mCrypton | 128 | 64 | 0.35 | 2594 | 4.66 | 138.61 | 33.51 | 12.91 | 96 | 64 | 1076 | 28 | 16457 | 68 | 15.5 | 14.41 |
| NOEKEON | 128 | 128 | 0.35 | 2604 | 4.68 | 1362.21 | 3.44 | 1.32 | 128 | 128 | 364 | 32 | 23517 | 95.9 | 21.7 | 59.62 |
| ICEBERG | 128 | 64 | 0.18 | 5817 | 8.72 | 21.81 | 400 | 68.76 | - | - | - | - | - | - | - | - |
| PUFFIN-2 | 80 | 64 | 0.18 | 1083 | 1.62 | 314.74 | 5.2 | 4.8 | - | - | - | - | - | - | - | - |
| PRINCE | 128 | 64 | 0.13 | 2953 | 2.95 | 5.53 | 533.3 | 180.59 | 128 | 64 | 1108 | 0 | 3614 | 14.4 | 70.8 | 63.9 |
| PRIDE | - | - | - | - | - | - | - | - | 128 | 64 | 266 | 0 | 1514 | 6 | 169 | 635.34 |
| PRINT | 80 | 48 | 0.18 | 503 | 0.75 | 7.54 | 100 | 198.8 | 80 | 64 | 1268 | 18 | 6095 | 25.1 | 42 | 33.12 |
| Klein | 64 | 64 | 0.18 | 1220 | 1.83 | 59.18 | 30.9 | 25.32 | 80 | 64 | 2164 | 368 | 35161 | - | 7.28 | 3.36 |
| LED | 64 | 64 | 0.18 | 966 | 1.45 | 282.55 | 5.1 | 5.27 | - | - | - | - | - | - | - | - |
| I-PRESENT | 80 | 64 | 0.18 | 2467 | 370 | - | - | - | - | - | - | - | - | - | - | - |
| EPCBC | 96 | 48 | 0.18 | 1008 | 1.51 | 124.74 | 12.12 | 12.02 | - | - | - | - | - | - | - | - |
| DESL | 56 | 64 | 0.18 | 1848 | 2.77 | 44.4 | 44.02 | 24.02 | 56 | 64 | 3098 | 0 | 8365 | 33.4 | 30.6 | 9.88 |
| TEA | 128 | 64 | 0.18 | 2355 | 3.53 | 35.32 | 100 | 42.46 | - | - | - | - | - | - | - | - |
| XTEA | - | - | - | - | - | - | - | - | 128 | 64 | 504 | 0 | 17514 | 70 | 14.6 | 28.97 |
| Camellia | 128 | 128 | 0.18 | 6511 | 9.76 | 33.57 | 290.1 | 44.55 | 128 | 128 | 1262 | 12 | 64000 | 256 | 8 | 6.34 |
| SIMON | 96 | 48 | 0.13 | 763 | 0.76 | 48.32 | 15.8 | 20.7 | 96 | 48 | 170 | 0 | 594 | 2.3 | 323 | 1900 |
| SEA | 96 | 8 | 0.13 | 2562 | 2.56 | 1117.67 | 2.29 | 0.89 | 96 | 96 | 426 | 24 | 41604 | 173.7 | 9.2 | 21.6 |
| KASUMI | 128 | 64 | 0.13 | 3437 | 3.44 | 29.9 | 115.14 | 33.5 | 128 | 64 | 1264 | 24 | 11939 | 47.6 | 21.4 | 16.93 |
| MIBS | 64 | 64 | 0.18 | 1396 | 2.09 | 10.47 | 200 | 143.26 | 64 | 64 | 3184 | 29 | 49056 | 66.2 | 5.2 | 1.63 |
| LBlock | 80 | 64 | 0.18 | 1320 | 2 | 9.9 | 200 | 151.51 | 80 | 64 | 976 | 58 | 18988 | 25.6 | 13.48 | 13.81 |
| ITUbee | - | - | - | - | - | - | - | - | 80 | 80 | 716 | 0 | 2607 | 10.4 | 122.7 | 171.37 |
| GOST | 256 | 64 | 0.18 | 1000 | 1.5 | 7.5 | 200 | 200 | 256 | 64 | 4748 | 190 | 10240 | 13.8 | 25 | 5.27 |
| Robin | - | - | - | - | - | - | - | - | 128 | 128 | 1942 | 80 | 4935 | 6.6 | 103.74 | 53.42 |
| Fantomas | - | - | - | - | - | - | - | - | 128 | 128 | 1920 | 78 | 3646 | 4.9 | 140.42 | 73.14 |
| CLEFIA | 128 | 128 | 0.13 | 2678 | 2.67 | 36.82 | 76 | 28.37 | 128 | 128 | 3046 | 0 | 28648 | 114.5 | 17.8 | 5.84 |
| PICCOLO | 80 | 64 | 0.13 | 1136 | 1.13 | 4.8 | 237.04 | 208.66 | 80 | 64 | 966 | 70 | 21448 | 28.9 | 11.93 | 12.35 |
| TWINE | 80 | 64 | 0.09 | 1503 | 1.05 | 5.91 | 178 | 118.42 | 80 | 64 | 1180 | 140 | 20505 | - | 12.48 | 10.58 |
| SPECK | 96 | 48 | 0.13 | 884 | 0.88 | 73.67 | 12 | 13.57 | 96 | 48 | 134 | 0 | 408 | 1.6 | 470.5 | 3511.19 |
| IDEA | - | - | - | - | - | - | - | - | 128 | 64 | 596 | 0 | 2700 | 10.8 | 94.8 | 159.06 |
| HIGHT | 128 | 64 | 0.35 | 2608 | 4.7 | 24.93 | 188 | 72.08 | 128 | 64 | 5718 | 47 | 6377 | 25.5 | 40.14 | 7.02 |
| LEA | 128 | 128 | 0.13 | 3826 | 3.82 | 50.22 | 76.19 | 19.91 | 128 | 128 | 590 | 32 | 5231 | - | 97.8 | 165.76 |
| KATAN | 80 | 32 | 0.13 | 802 | 0.8 | 64.16 | 12.5 | 15.58 | 80 | 64 | 338 | 18 | 72063 | 289.2 | 3.5 | 10.36 |
| KTANTAN | 80 | 32 | 0.13 | 462 | 0.46 | 36.96 | 12.5 | 27.05 | 80 | 32 | 10516 | 614 | 10233211 | 13814.8 | 0.012 | 0 |
| Hummingbird | - | - | - | - | - | - | - | - | 128 | 16 | 1822 | 82 | 4637 | 6.2 | 13.8 | 7.57 |
| Hummingbird-2 | 128 | 16 | 0.18 | 2159 | 3.23 | 40.48 | 80 | 37.05 | 128 | 16 | 770 | 50 | 1520 | 2 | 42.1 | 54.68 |

**Table 3.** Hardware and Software performances of LWC algorithms

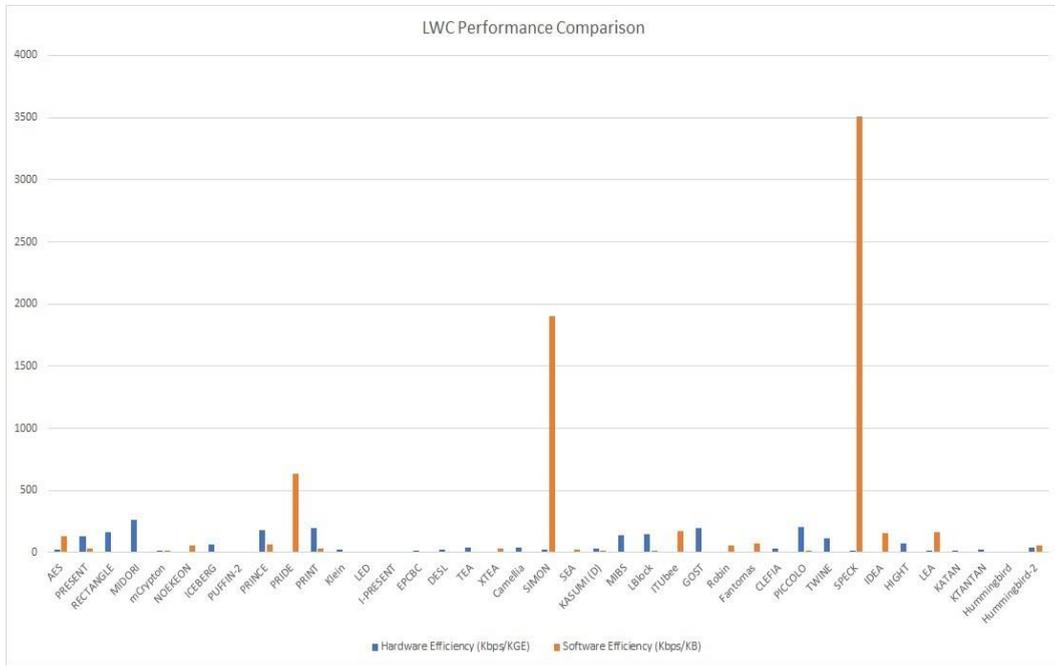

**Figure 3.** LWC Performance Comparison

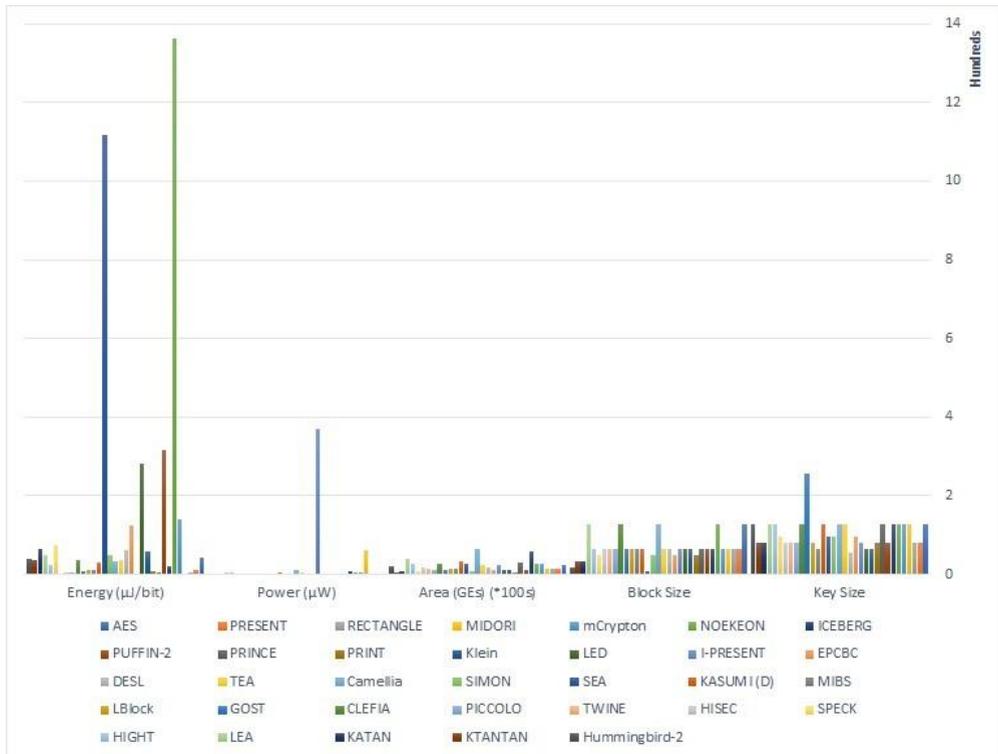

**Figure 4.** Comparison by Hardware Metrics

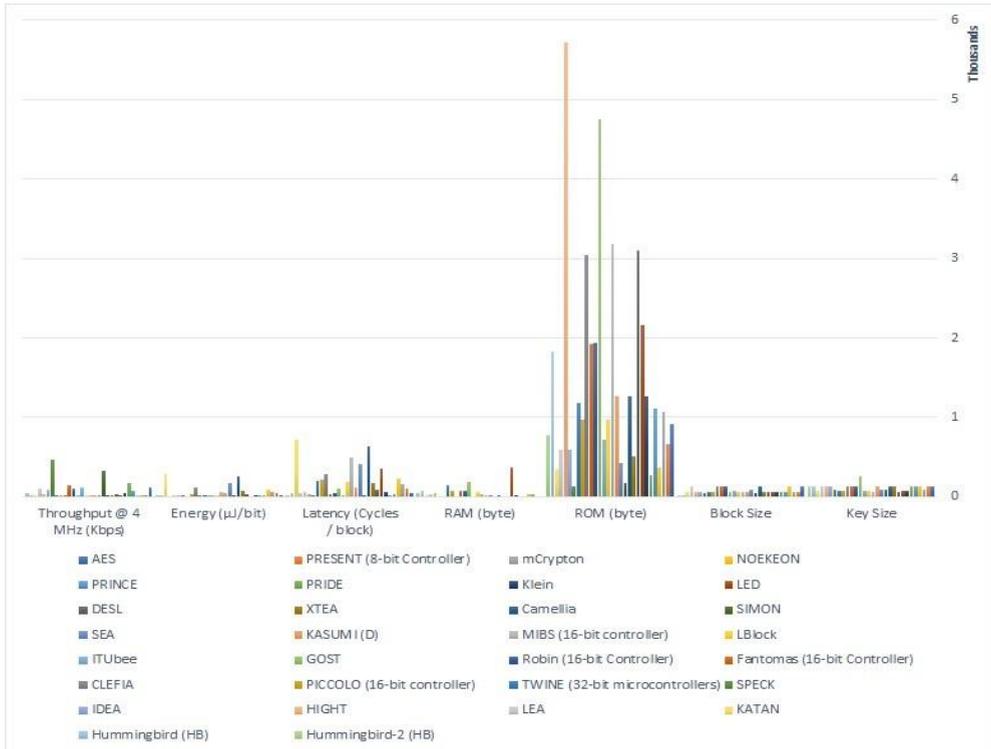

**Figure 5.** Comparison by Software Metrics

## V. CRYPTANALYSIS OF LWC ALGORITHMS

Along with performance and cost, security is an important and an essential measure for any lightweight cryptography algorithm. Attack resistance property of any lightweight cryptography algorithm can be measured by means of crypt- analysis. Cryptanalysis is aimed to find weaknesses of the algorithm and developing a method of decryption [34]. There are four main types of attacks on block cipher [34] [48] [49] [98]: Differential cryptanalysis, Linear cryptanalysis, Integral cryptanalysis and Algebraic attacks. This cryptanalysis attacks are based on Ciphertext only, Known plaintext, Chosen plaintext and Chosen ciphertext along with Man-in-the- middle attack, Brute force and Side channel attacks. Table 4 highlights the possible attacks on various lightweight cryptography algorithms [36]. Almost all the existing lightweight block cipher solutions suffers from related key attack, followed by the differential cryptanalysis. Moreover, the lighter version (with reduced rounds) are more vulnerable to various attacks compared to their standard one.

### A. STANDARDIZATION OF LWC ALGORITHMS

NIST (National Institute of Standards and Technology), ISO/IEC (International Organization of Standardization and the International Electrotechnical Commission), Cryptrec (Cryptography Research and Evaluation Committees (set up by Japanese Government)), Ecrypt (European Network of Excellence in Cryptology), NSA (National Security Agency of USA) and CryptoLUX (University of Luxembourg) are some of the organizations/research groups, who are actively contributing in the field of cryptography to improve the lightweight standards for resource constrained devices.

PRESENT [48] and CLEFIA [115] are the only two algorithms approved by the ISO/IEC 29192 standard whereas AES, Camellia, CLEFIA, TDES, LED, PRESENT, PRINCE, Piccolo, TWINE, SIMON and SPECK, Midori are targeted by Cryptrec.

Based on the above standardizing bodies and popular use in different applications by resource constrained devices such as embedded systems, wireless sensors and IoT de- vices, our targeted algorithms are AES, PRESENT, RECT- ANGLE, I-PRESENT, DESL/DESXL, CLEFIA, HIGHT, SIMON and SPECK, TWINE,

TEA, Piccolo, Midori, KATAN/KTANTAN, and Hummingbird-2 where they use either SP or FN structure (or mix of both).

## B. VARIOUS APPLICATIONS & THEIR DEMANDS OF LWC ALGORITHMS

The wide range of IoT applications in various fields creates the demand for lightweight cryptography algorithms in different circumstances. Some of the emerging applications are listed in Table 5 [154]:

| Sr. No. | Algorithm | Cryptanalysis |
|---|---|---|
| 1 | AES | Biclique Cryptanalysis [149], Boomerang [47], related key attack [47] |
| 2 | PRESENT | Side-channel attacks [150] [151], related key attack on the 17-round [132], improved differential fault analysis [152], Biclique-attacks on full round [24], truncated differential attack on 26-round [153] |
| 3 | RECTANGLE | Side-channel attack [49], related-key cryptanalysis [49], statistical saturation attack [49] |
| 4 | MIDORI | Differential and Linear Cryptanalysis on reduced rounds, Boomerange, impossible differential attack on 6-round, Meet-in-the-Middle Attacks, brute force attacks [37] |
| 5 | mCrypton | A related key rectangle attack for 8-round [53] |
| 6 | NOEKEON | Related-key cryptanalysis [55] |
| 7 | ICEBERG | Differential cryptanalysis on 8-round [58] |
| 8 | PUFFIN-2 | Differential cryptanalysis [61] |
| 9 | PRINCE | Attacks on the reduced 6-round and the full 12-round [65] [66], truncated differential cryptanalysis on 7-round [67] |
| 10 | PRINT | Related-key cryptanalysis [69] |
| 11 | Klein | Chosen-plaintext key-recovery attacks [71], an asymmetric biclique attack [72] |
| 12 | LED | related key attacks [75], Biclique cryptanalysis on reduced round cipher [24], differential fault analysis based on Super-Sbox techniques [76] |
| 13 | ZORRO | Practical invariant subspace attacks [79] |
| 14 | EPCBC | Algebraic attack up to 5 rounds of EPCBC-96 and up to 8 rounds of EPCBC-48 using known plaintexts, weak keys for EPCBC-96 with up to 7 rounds [19] |
| 15 | TEA | Simple key scheduling vulnerable to brute force attack [88] [89] |
| 16 | XTEA | Related-key rectangle attack on 36 rounds [91] |
| 17 | XXTEA | Chosen-plaintext attack based on differential analysis against the full-round cipher [93] |
| 18 | Camellia | Cache timing attacks [20], Impossible differential attack [95] |
| 19 | SIMON | Differential and impossible differential attacks [25], differential fault attack [97], Cube and dynamic cube attacks [26], recover the full key in a practical time complexity [26], many on reduced round version [98] |
| 20 | KASUMI | Differential-based related-key attack [13], single-key attack on 6-round [14] |
| 21 | MIBS | Linear attacks up to 18 rounds, first ciphertext-only attacks on 13 rounds, differential analysis up to 14 rounds and impossible differential attacks up to 12 rounds [105] |
| 22 | LBlock | Saturation attack on 22-round [107], biclique cryptanalysis against the full round [108], improved differential fault attacks [21] and round addition differential fault analysis [22] |
| 23 | ITUbee | Self-similarity cryptanalysis on 8-round [113] |
| 24 | GOST | Theoretical meet-in-the-middle attack [110], a single-key attack [111] |
| 25 | Robin | Practical invariant subspace attacks [79] |
| 26 | CLEFIA | Key recovery attach on 10-round [116], saturation cryptanalysis [116] |
| 27 | PICCOLO | Biclique attacks on full round [24] [120], impossible differential cryptanalysis on reduced round versions [121] |
| 28 | TWIS | Differential distinguisher with probability 1 [123] |
| 29 | TWINE | Meet-in-the-middle attacks on simplified key scheduling [124], saturation attack [84] |
| 30 | SPECK | Various attacks on reduced versions [126], differential fault attacks [97], Key recovery [98], boomerange attack [98] |
| 31 | IDEA | Narrow-biclique attacks [12] |
| 32 | HIGHT | Impossible differential attack on 26-round [132], a related-key attack on full round [133], biclique cryptanalysis on full round version [120] and zero-correlation attacks on the 26 and 27-round cipher [11] |
| 34 | LEA | Power analysis (retrieving 128-bit key) [23] |
| 35 | KeeLoq | Slide attack, novel meet-in-the-middle attack (direct algebraic attacks can break up to 128 rounds of KeeLoq) [138] [139] [137] |
| 36 | KATAN | Multi-dimensional meet-in-the-middle attacks on reduced round [141] |
| 37 | KTANTAN | Related-key attacks on full 80-bit key [15] |
| 38 | Hummingbird | Vulnerable to several attacks [144] |
| 39 | Hummingbird-2 | Related key attack [146] |

**Table 4.** Possible Attacks on LWC Algorithms

| Applications | Key Requirements | Effective Algorithm(s) |
|---|---|---|
| Electrical Home Appliances (Smart TV, Fridge, etc.) | Less CPU Time, Smaller ROM | SIMON, SPECK, Piccolo, TWINE |
| Logistics (RFID tag) | Small circuit size, Low power | SIMON, SPECK, Piccolo, PRINCE |
| Smart Agriculture (Sensors) | Compact implementation, Less processing cycles, Low power consumption | SIMON, SPECK, PRESENT, TWINE |
| Medical (Sensors) | Low power consumption, low CPU cost, Small circuit size | SIMON, SPECK, PICCOLO, PRESENT |
| Industrial systems | Real-time processing | AES |
| Automobiles | Small circuit size, lower latency | Keeloq, Midori, PRINCE, PRESENT, SIMON |

**Table 5.** Applications & their demands of LWC algorithms

## VI. OPEN RESEARCH CHALLENGES AND RESEARCH DIRECTIONS

The ideal algorithm should maintain a proper balance among cost, performance and security (Figure 6). Any two of the three design goals, security and low costs, security and performance, low costs and performance can be easily optimized, whereas it is very difficult to optimize all three design goals at the same time [34]. For example, block ciphers, the key length provides a security-cost trade-off, while the number of rounds provides a security-performance trade-off and the hardware architecture a cost-performance trade-off [34] (increasing number of rounds [98] or key size results in degradation of algorithm performance). These could be achieved by design focus on less memory and less computing power requirement, leading to less Gate Equivalent (physical area) requirements along with low power (energy) consumption without compromising strong security [31]. Based on the above study, we have identified the following research issues, which require further attention to make the LWCs algorithms effective in IoT security:

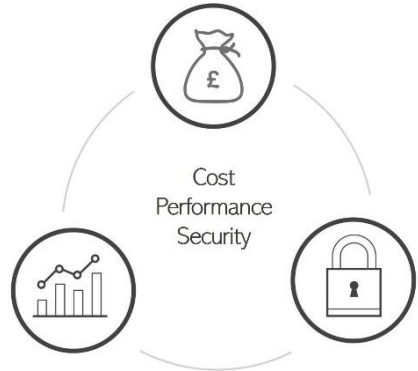

**Figure 6.** Cost, Performance and Security

1) An efficient masking for a block cipher comes mainly from the S-boxes (by choosing an adequate S-box). In addition, S-box should demonstrate a good trade-off between efficiency, conventional security and masking efficiency [77]. So designing simple and fast but strong confusion (Substitution, S-box) and diffusion (Bit Per- mutation) properties with right balance amongst cost, performance and security is of practical interest, e.g., How to reduce number of S-boxes as they increase the demands for memory (to store) and computing power (to produce) while maintaining the same security level? (motivation: PRESENT is designed from AES and replaces eight S-boxes with just one. Similarly, many researchers have derived the lighter versions from the standard cryptography algorithms with a few modifications by reducing substitution-permutation (counter- effect on security level)). But how to replace S-boxes with some other confusion techniques with the same level of security and less overhead of memory and processing cost is still an open problem.
2) Making key scheduling lighter with smaller key size and adequate strength, i.e., How to generate random sub-keys from the provided initial key for all n rounds? Different sub-keys (for all n rounds) should be generated from the initial key for every attempts.

3) Increase in number of rounds adversely affects the performance and cost, i.e., How to decrease (or increase) number of rounds without compromising performance as well as security level?

We are currently working on substitution-permutation net- work (SPN) methods with main focus on S-Box to design a generic lightweight cryptography algorithm with right blend of three main characteristics namely, cost, performance and security.

## VII. CONCLUSION

Due to the exponential growth in the number of IoT de- vices in various domains, IoT security is one of the main concerns and as a consequence, there is need of lightweight algorithm(s) with strong security and right balance of cost and performance metrics. For resource constrained devices, in particular IoT devices, lightweight cryptography is an effective way to secure the communication by transforming the data. The well-defined LWC characteristics (cost, performance and security) by NIST are compared and further research gaps and open research challenges are highlighted in this paper. From the literature review, AES, PRESENT, CLEFIA and lightweight versions of DES are the most experimented and widely accepted block ciphers. However, new attacks are reported with the growth of new LWC algorithms which is inevitable and never ending process. The war between cyber security experts and attackers always opens a door of opportunities for new research in the field of cybersecurity, especially lightweight cryptography.